# Dynamic competition between spin density wave order and superconductivity in underdoped $Ba_{1-x}K_xFe_2As_2$

M. Yi[1,2], Y. Zhang[1,3], Z.-K. Liu[1,2], X. Ding[4], J.-H. Chu[5], A. F. Kemper[6], N. Plonka[1,2], B. Moritz[1,7], M. Hashimoto[8], S.-K. Mo[3], Z. Hussain[3], T. P. Devereaux[1], I. R. Fisher[1,2], H.H. Wen[4], Z.-X. Shen[1,2*], and D. H. Lu[8*]

[1]*Stanford Institute for Materials and Energy Sciences, SLAC National Accelerator Laboratory and Stanford University, Menlo Park, California 94025, USA*

[2]*Departments of Physics and Applied Physics, and Geballe Laboratory for Advanced Materials, Stanford University, Stanford, California 94305, USA*

[3]*Advanced Light Source, Lawrence Berkeley National Lab, Berkeley, California 94720, USA*

[4]*Center for Superconducting Physics and Materials, National Laboratory of Solid State Microstructures and Department of Physics, National Center of Microstructures and Quantum Manipulation, Nanjing University, Nanjing 210093, China*

[5]*Department of Physics, University of California, Berkeley, California 94720, USA*

[6]*Computational Research Division, Lawrence Berkeley National Lab, Berkeley, California 94720, USA*

[7]*Department of Physics and Astrophysics, University of North Dakota, Grand Forks, North Dakota 58202, USA*

[8]*Stanford Synchrotron Radiation Lightsource, SLAC National Accelerator Laboratory, Menlo Park, California 94025, USA*

*To whom correspondence should be addressed: dhlu@slac.stanford.edu and zxshen@stanford.edu

**An intriguing aspect of unconventional superconductivity is that it always appears in the vicinity of other competing phases, whose suppression brings the full emergence of superconductivity. In the iron-pnictides, these competing phases are marked by a tetragonal-to-orthorhombic structural transition and a collinear spin-density-wave (SDW) transition. There has been macroscopic evidence for competition between these phases and superconductivity as the magnitude of both the orthorhombicity and magnetic moment are suppressed in the superconducting state. Here, using angle-resolved photoemission spectroscopy on detwinned underdoped $Ba_{1-x}K_xFe_2As_2$, we observe a coexistence of both the SDW gap and superconducting gap in the same electronic structure. Furthermore, our data reveal that following the onset of superconductivity, the SDW gap decreases in magnitude and shifts in a direction consistent with a reduction of the orbital anisotropy. This observation provides direct spectroscopic evidence for the dynamic competition between superconductivity and both SDW and electronic nematic orders in these materials.**

**Introduction**

To understand the phenomenon of high temperature superconductivity (HTSC), it is important to understand the competing states of the superconducting phase. In both families of HTSC materials, the competing states consist of various symmetry-breaking phases, which interact with superconductivity (SC) in intricate ways. In the cuprates, one of the biggest puzzles is the nature of the pseudogap (PG) phase, which envelops a large portion of the superconducting dome in the phase diagram. Recently, a large amount of experimental evidence suggests a competition between the PG and SC[1-4]. However, the controversial nature of the PG phase itself complicates a direct study of this competition. Conversely, in the iron pnictides, the competing phases are well characterized: a collinear spin-density wave (SDW) order has been well established as the main competing phase in the underdoped region of iron-pnictides[5-6], while a nematic phase associated with the structure transition ($T_S$) has been suggested as another competing phase[7-11]. The two transitions associated with these competing orders track each other closely, occurring either simultaneously or with $T_S$ preceding $T_{SDW}$. SC emerges as these phases are suppressed with carrier doping, isovalent substitution, or pressure[12-13]. The way in which these two orders interact with SC varies among the pnictide families: while SC and SDW exhibit mutual exclusion in the $Ln$FeAsO families[14], there is a large region of coexistence in both the NaFeAs[15] and the BaFe$_2$As$_2$ families[16-18]. In addition, transport and neutron scattering experiments have shown that the orthorhombicity and magnetic moment, which are the macroscopic order parameters for the structural and SDW orders, are suppressed with the onset of superconductivity[19-21], signaling macroscopic competition between these orders and superconductivity.

In this paper, we focus on the phase competition between these orders and SC, which makes underdoped Ba$_{1-x}$K$_x$Fe$_2$As$_2$ in the coexistence region an ideal candidate as $T_S$/$T_{SDW}$ and $T_C$ are well separated yet $T_C$ is sufficiently high to be experimentally accessible[18]. In particular, we study the electronic structure of an underdoped (UD) compound (x = 0.15), where $T_S$ = $T_{SDW}$ = 110K and $T_C$ = 14.5K, and the non-superconducting undoped BaFe$_2$As$_2$ compound ($x$ = 0) with $T_S$ = $T_{SDW}$ = 138K for comparison. Using high-resolution angle-resolved photoemission spectroscopy (ARPES), we directly observe the electronic response of each order in this compound. Distinct spectral signatures of the three orders are observed to coexist on the same electronic structure below $T_C$, signaling the microscopic coexistence of the three orders. More importantly, we observe the spectral gap associated with the SDW order to be suppressed with the onset of SC. Furthermore, using a tight-binding model, we show that a reduction in the nematic order parameter is also needed to fully account for the observed shift of the SDW gap below $T_C$. These observations show a direct electronic competition between the SDW and nematic orders with superconductivity in $Ba_{1-x}K_xFe_2As_2$.

**Results**

Spectral Evolution Across Phase Transitions

The momentum-resolved electronic structure, which can be directly imaged by ARPES, is expected to undergo distinct changes at each of the transitions in the system. As illustrated in the generic schematic in Fig. 1b, starting in the normal state ($T > T_S, T_{SDW}$), the typical electronic structure of iron-pnictides consists of hole bands at the Brillouin zone (BZ) center, Γ point, and electron bands at the BZ corner, X point. Lowering the temperature approaching $T_S$, orbital-dependent band shifts onset, as has been observed in Co-doped $BaFe_2As_2$[9] and NaFeAs[22], where bands dominated by the $d_{xz}$ ($d_{yz}$) orbital character shift down (up) in a $k$-dependent fashion (minimal at Γ and maximal at X, see Supplementary Figure 1, Supplementary Note 1), breaking $C_4$ rotational symmetry. Across $T_{SDW}$, translational symmetry is broken and bands fold along $\mathbf{Q_{SDW}} = (\pi,\pi)$, opening SDW gaps such that the Fermi surface (FS) reconstructs, leaving portions ungapped. Finally, as the system is cooled below $T_C$, particle-hole symmetric superconducting gaps open on bands that cross the Fermi level, $E_F$. Hence, there are distinct spectroscopic signatures associated with the three order parameters in this system.

To avoid twinning effects in the orthorhombic state, all measurements were performed on mechanically detwinned samples held under uniaxial stress. Although the effect of the detwinning stress on the transition temperatures has not been systematically studied in the $Ba_{1-}$ $_x$K$_x$Fe$_2$As$_2$ system, such study in the Ba(Fe$_{1-x}$Co$_x$)$_2$As$_2$ system shows that the stress naturally broadens $T_S$ to higher temperature[7], but does not affect the full magnitude of the orbital anisotropy at the lowest temperature[9] (Supplementary Note 1). Furthermore, in a systematic study in the BaFe$_2$(As$_{1-x}$P$_x$)$_2$ system, it is shown that the mechanical detwinning stress increases $T_{SDW}$ and decreases $T_C$[23]. In the current study, we do not observe any significant modification of $T_{SDW}$ and $T_C$ within experimental uncertainties inferred from the temperatures at which spectral features associated with these orders onset compared with those measured on unstressed crystals.

Figure 1c shows the changes observed across $T_S$/$T_{SDW}$ for the UD sample. In the normal state, the Γ-X and Γ-Y directions are degenerate, respecting C$_4$ symmetry. Lowering temperature below $T_S$/$T_{SDW}$ lifts this degeneracy, leading to the observed C$_4$ symmetry breaking. More specifically, measuring band dispersion with light polarized perpendicular to the cut direction (Fig. 1d upper panels) reveals that the hole-like band with predominantly d$_{yz}$ character along the antiferromagnetic direction near X moves to higher energy than the corresponding d$_{xz}$ band along the ferromagnetic direction near Y, a signature of the orbital anisotropy seen in the nematic phase of Co-doped BaFe$_2$As$_2$[9]. In fact, the magnitudes of the orbital energy splitting in the K-doped samples are comparable with those in Co-doped samples, and roughly scale with $T_S$ independent of the type of doping (Supplementary Figure 2). Furthermore, electron bands folded from the BZ corner can be observed at the Γ point, indicative of long range SDW order. These signatures of the nematic and SDW orders in the UD case are very similar to those observed in BaFe$_2$As$_2$, as a chemical potential shift of ~50meV roughly mimics the doping effect between BaFe$_2$As$_2$ and the UD sample (Supplementary Note 1).

SDW gaps open where the folded electron bands cross the hole bands near Γ (Fig. 2). In BaFe$_2$As$_2$, the SDW gaps can be seen around -60meV and -15meV along those cuts (Fig. 2a).

Focusing on the deeper SDW gap, the gap opens at $T_{SDW}$ and monotonically increases with lowering temperature as no other transitions exist below $T_{SDW}$, similarly to the behavior of the magnetic moment[20-21]. The more interesting case occurs in the UD compound, in which an SDW gap around -15meV corresponds to the one discussed in BaFe$_2$As$_2$. Figures 2c-d compare the evolution of this SDW gap above (20K) and below (5K) $T_C$, where the bands forming the gap are marked. Cooling the sample below T$_C$, the SDW gap magnitude abruptly decreases, which can be better visualized in two ways. First, comparing the energy distribution curves (EDCs) above and below $T_C$ (Fig. 2e), we see two changes: i) at $k_F$ where the upper band crosses $E_F$, a SC gap opens (circles); ii) the lower band shifts up in energy (dashed curves), reducing the effective SDW gap size. Secondly, we can examine the momentum distribution curves (MDCs), which are horizontal line cuts through the spectra images in Fig. 2c-d. We fit these MDCs with Lorentzians for the two hole bands (Supplementary Note 2). As the bands are relatively steep, the fitted positions would trace out the underlying ungapped band dispersion, while the fitted intensity of the inner hole band would show a suppression within the energy range in which the SDW gap develops. The fitted intensity for the inner hole band is plotted in Fig. 2g for the two temperatures, where the separation between the two peaks is another measurement of the SDW gap, which clearly decreases below $T_C$.

Model Simulation

To better understand this decrease in the SDW gap below $T_C$, we have simulated the spectroscopic effects of three order parameters ($\Delta_S$, $\Delta_{SDW}$, $\Delta_{SC}$) associated with the three transitions ($T_S$, $T_{SDW}$, $T_C$) based on a five-orbital tight binding model of BaFe$_2$As$_2$ from DFT band structure[24] which is shifted by 0.15eV and renormalized by a factor of 3 to match the observed band structure for the UD compound. The order parameters are incorporated phenomenologically, with $\Delta_S$ introduced as an orbital-dependent on-site energy shift between $d_{xz}/d_{yz}$ that is momentum-dependent (zero at $\Gamma$ and maximal at X/Y); $\Delta_{SDW}$ as an interaction term in the orbital-basis; and $\Delta_{SC}$ as a particle-hole symmetric isotropic gap at $E_F$. We examine the calculated band dispersions near $\Gamma$ along the same cut as the data in Fig. 2c. Figure 3a displays the situation deep in the orthorhombic SDW state but still above $T_C$, with finite values for $\Delta_S$ and $\Delta_{SDW}$. Here we highlight in pink the SDW gap formed between the inner hole band and the folded electron band similar to the one observed experimentally in Fig. 2c. In the simplest scenario with no competition and superconductivity added directly (Fig. 3b), only bands near $E_F$ are affected by the opening of a SC gap, with little effect on the observed SDW gap. In particular, the lower branch does not shift up, contrary to experimental observation. To account for the observed reduction in the SDW gap below $T_C$, $\Delta_{SDW}$ needs to be reduced before adding $\Delta_{SC}$ (Fig. 3c). However, the center of the gap does not shift noticeably relative to that above $T_C$, whereas in the data, an overall shift towards $E_F$ is also observed in addition to the reduced gap size. To reproduce this shift, an additional reduction in $\Delta_S$ is required (Fig. 3d). The best match to the experimental data is obtained by introducing a moderate reduction in both $\Delta_{SDW}$ and $\Delta_S$ (Fig.3e). In reality, these two order parameters are intimately connected, as one enhances the other[25-27]. It is intrinsically difficult to separate these two effects, but the exercise displayed in Fig. 3 demonstrates the qualitative effects on the electronic structure of reducing each order parameter, and gives an intuitive understanding of the combined suppression of both orders below $T_C$. Here we note that the observed SDW gap shrinks in magnitude, as seen in the shift of the EDC peaks marking the boundary of the gap. This is the behavior expected of the spectral evolution of a generic order parameter in competition. We note that our theoretical model here is based on an itinerant origin of the magnetism, while the general nature of magnetism in iron pnictides has not been fully settled, with proposals for coexistence of both itinerant and local magnetism[28]. In this regard, our observation of the reduction of the SDW gap itself shows a general competition of SC and SDW independent of the nature of the magnetism, as the gap size reflects the order parameter. The modeling presented here serves to demonstrate the effects of the competition on the electronic spectra.

Detailed Temperature Dependence

Finally, we can trace the temperature-dependent evolution of this SDW gap by tracking the EDCs across the SDW gap (Fig. 4a), where the separation between the two peaks at each temperature represents the effective SDW gap magnitude. The SDW gap forms below $T_{SDW}$, but instead of monotonically increasing with decreasing temperature as in $BaFe_2As_2$, it dips at the onset of SC (Fig. 4c). This behavior is reminiscent of that of the orthorhombicity and magnetic moment observed in UD compounds[20-21]. To complete this comparison, we also plot the temperature dependence of the SC gap on the same inner hole band, which onsets at $T_C$ (Fig. 4b-c). Here we note that the observed reduction in the SDW gap (~30%) seems much larger than that in the moment (~4%) for this doping[20-21]. The reason is that this observed decrease in the gap is actually a combined result of three effects: i) opening of the SC gap pushing down the upper branch; ii) reduction of $\Delta_{SDW}$; and iii) reduction of $\Delta_S$, which occur constructively at this particular momentum to magnify the effect. Depending on the precise interplay of these three ingredients, the overall effect at other momenta could be smaller, but the observed reduction in SDW gap is not limited to this momentum (Supplementary Note 3). To further confirm such behavior, we have performed mean-field calculations using tight binding fits to the DFT derived bands, including an ad-hoc isotropic SC order parameter and self-consistently determined orbital-dependent SDW. The results show that the introduction of SC indeed suppresses the SDW gaps, as naively expected from the usual interplay between SC and magnetism, where SC gaps quasiparticles around the Fermi surface, leaving much less spectral weight to simultaneously form the SDW order. The temperature dependence of the $\Delta_{SDW}$ follows a trend similar to that shown in Fig. 4c (Supplementary Note 4).

**Discussion**

The coexistence of magnetism and superconductivity holds a special fascination in the field of high temperature superconductivity as magnetism is generally believed to be detrimental to SC[29-30]. Extensive efforts have been made to understand whether this coexistence in the pnictides is indeed microscopic as it would have strong implications on the mechanism of SC in these compounds. In particular, theoretical work has shown that a conventional $s_{++}$ pairing symmetry state cannot coexist with the SDW order whereas such coexistence is possible for an unconventional $s_{\pm}$ pairing symmetry[31]. Using a simple simulation of the APRES spectra, we show that our observation of the reduced SDW gap and its shift towards $E_F$ as well as the observation of a well-defined superconducting gap below $T_C$ on the same spectrum cannot be explained by a phase separated coexistence of these orders but more consistent with a microscopic coexistence scenario (Supplementary Note 5), in agreement with NMR measurements[32], pointing to $s_{\pm}$ pairing symmetry for this part of the phase diagram. Furthermore, it is clear from our results that the SDW and nematic order are not mere spectators to the presence of SC. The SDW gap—a representation of the strength of the coherent electronic excitations of the SDW order—is suppressed with the onset of SC, indicating that the SDW order competes with SC for the same electrons. The fact that a reduction in both $\Delta_{SDW}$ and $\Delta_S$ is required to fully account for the experimental observations also demonstrates that the SDW order and the electronic nematicity are intimately connected and both are likely in competition with SC in the coexistence region.

Lastly, it is interesting to point out the connection to the cuprates. Traditionally, there have been two main interpretations of the PG phase: i) a precursor state of SC, ii) a competing phase to SC. While several recent reports favour the scenario of competition[1-4], the elusive nature of the PG complicates the interpretations. Here, we demonstrate clearly with spectroscopic evidence the competition between SC and a well-characterized long-range density wave, which is intriguingly similar to a recent study in cuprates showing a similar suppression of the PG spectral features associated with the onset of SC (M. Hashimoto et al. submitted). While the competition in iron pnictides may be simpler than that in the cuprates, the comparison of the two families may provide helpful insights on the mysterious PG phase in the cuprates, and help link the two families for an eventual understanding of the mechanism of unconventional SC.

**Methods**

Experimental Conditions

High quality single crystals of $Ba_{1-x}K_xFe_2As_2$ were grown using the self flux method[18]. Mechanical detwinning of the crystals was performed using a previously reported method[9]. ARPES measurements were carried out at both beamline 5-4 of the Stanford Synchrotron Radiation Lightsource and beamline 10.0.1 of the Advanced Light Source using SCIENTA R4000 electron analyzers. The total energy resolution was set to 5 meV and the angular resolution was 0.3°. Single crystals were cleaved in situ below 10 K for each measurement. All measurements were done in ultra high vacuum with a base pressure lower than $4\times10^{-11}$ torr.

Tight-Binding Model Simulations

The simulations used in Fig. 3 were produced based on a five-orbital tight binding model of BaFe$_2$As$_2$ from DFT band structure[24]. The $E_F$ was first shifted down by 0.15eV to account for the hole doping. The bands were then renormalized by a factor of 3 to match the observed band structure for the UD compound. The cut of the calculations shown in Fig. 3 was along $k_y = 0.07\pi/a$, to match the experimental cut shown in Fig. 2. The parameters used in Fig. 3 were implemented as described in the following. The orbital anisotropy, $\Delta_S$, is an onsite energy splitting between d$_{xz}$ and d$_{yz}$ that is momentum-dependent ($\Delta_S(\mathbf{k}) = \alpha \cdot \sin^2((|k_x|+|k_y|)/2)$, which is zero at $\Gamma$ and maximum at X/Y along the high symmetry directions, and also periodic in $k$, see Supplementary Note 1), where $\alpha\{d_{xz}, d_{yz}\} = \{-20\text{meV}, 40\text{meV}\}$. The SDW order parameter, $\Delta_{SDW}$, is diagonal in the orbital-basis, with $\Delta_{SDW}\{d_{xz}, d_{yz}, d_{x2-y2}, d_{xy}, d_{3z2-r2}\} = 30\text{meV}*\{0.25, 1, 0.25, 0.5, 0.25\}$. The ratios between the different orbitals are consistent with the results of self-consistent calculations for the SDW order in the orbital-basis, in which d$_{yz}$ bands tend to open the largest SDW gap[34]. The superconducting order parameter, $\Delta_{SC}$, is taken to be isotropic and 5meV.

**Acknowledgements** ARPES experiments were performed at the Stanford Synchrotron Radiation Lightsource and the Advanced Light Source, which are both operated by the Office of Basic Energy Sciences, U.S. Department of Energy. The Stanford work is supported by the US DOE, Office of Basic Energy Science, Division of Materials Science and Engineering, under Award # DE-AC02-76SF00515.

**Author Contributions** M.Y. and D.H.L. conceived the project. M.Y., Y.Z. and Z.L. took the measurements. M.Y. analyzed the data. X.D. grew the underdoped single crystals. J.H.C. grew the undoped single crystals. A.F.K., N.P., and B.M. performed the calculations and provided theoretical insights. M.H. and S.K.M. assisted in ARPES measurements at SSRL and ALS. Z.H., T.P.D., I.R.F., and H.H.W. provided guidance. Z.X.S. and D.H.L. oversaw the project. M.Y. and D.H.L. wrote the paper with input from all coauthors.

**Author Information** The authors declare no competing financial interests. Correspondence and request for materials should be addressed to D. H. Lu (dhlu@slac.stanford.edu) and Z.-X. Shen (zxshen@stanford.edu).

**Fig. 1: Schematic and spectral evolution across phase transitions**

(a) Phase diagram of $Ba_{1-x}K_xFe_2As_2$ adapted from Ref. [18], where the gray vertical line indicates the doping of the UD compound studied. (b) Schematic of the expected evolution of the band structure across $T_S$, $T_{SDW}$, and $T_C$. From left to right, the four panels exhibit the band structure along the high symmetry direction Γ-X in the 2-Fe Brillouin zone, starting with the normal state ($T>T_S$, $T_{SDW}$), with the incremental addition of orbital anisotropy ($T_S>T>T_{SDW}$), SDW folding and gapping ($T_S/T_{SDW}>T>T_C$), and finally superconductivity ($T<T_C$). (c) Measured electronic structure (Fermi surface and band dispersions) in the normal state and SDW state (below $T_S/T_{SDW}$ but above $T_C$) of the UD compound, with mixed even and odd light polarization. The reconstructed Fermi surface in the SDW state is outlined. (d) Band dispersions measured with purely even or odd polarization along the AFM and FM directions in the SDW state at 15K. The $k<0$ portions are second energy derivatives of the raw spectra. The hole-like band identified as $d_{yz}$ along AFM and $d_{xz}$ along FM are labeled, signaling the electronic nematicity. The electron bands folded from the X point to the Γ point are marked with yellow dotted curves. All measurements were taken with 47.5eV photons.

**Fig. 2: Suppression of SDW gap below $T_C$**

(a) Band dispersions of $BaFe_2As_2$ measured in the SDW state (10K) near Γ as labeled in (f). The left hand side is the raw image, the right hand side the second energy derivative. Guides to the eye for the dispersions are marked in color, highlighting the dominant orbital character (red for $d_{xz}$; green for $d_{yz}$). As the dispersions are the result of SDW folding and gapping, the inferred bands before SDW gapping are marked by yellow dashed lines. (b) Same as (a) but for the UD ($x=0.15$) sample, measured in the SDW state above $T_C$ at 20K. (c) The band dispersions from (b) near the SDW gap taken above $T_C$, divided by the Fermi function convolved with a Gaussian representing instrument resolution. Open symbols are fitted peak positions of the EDCs indicating the band dispersions. Blue arrow points to the SDW gap in discussion. (d) Same measurement but taken below $T_C$, at 5K. (e) Comparison of EDCs taken above $T_C$ (red, 20K) and below $T_C$ (blue, 5K) across the SDW gap in the momentum range marked by the black arrows in (c)-(d). Dashed lines trace out the band dispersions. Black circles highlight opening of the SC gap at $k_F$. (f) Schematic of the reconstructed FS in the SDW state. (g) The MDCs of (c)-(d) are fitted with two Lorentzians and a linear background. The intensity of the inner hole band is plotted above (red) and below (blue) $T_C$, where the separation of the two prominent peaks (black marks), measuring the effective SDW gap, shrinks below $T_C$. Measurements were taken with 25eV photons.

**Fig. 3: Model simulation for the three order parameters**

A tight-binding fit of BaFe$_2$As$_2$ to DFT band structure is used, with a shift of 0.15 eV of the chemical potential and a renormalization factor of 3 to match the measured dispersion of the UD compound. The three order parameters are introduced phenomenologically to understand the effect on spectroscopy: ($\Delta_S$, $\Delta_{SDW}$, $\Delta_{SC}$, associated with $T_S$, $T_{SDW}$, $T_C$). Calculations for the same cut as that in Fig. 2 are plotted here, with the SDW gap highlighted in pink. (a) Bands with only finite $\Delta_S$ and $\Delta_{SDW}$, mimicking the situation in the orthorhombic SDW state above $T_C$. (b) Simple addition of finite $\Delta_{SC}$ with no change in $\Delta_S$ and $\Delta_{SDW}$, opening a SC gap at $E_F$. The SDW gap is not much affected. (c) With addition of finite $\Delta_{SC}$ and a reduced $\Delta_{SDW}$, the SDW gap size is reduced, but the center does not shift. (d) With the addition of finite $\Delta_{SC}$ and reduced $\Delta_S$, the effective SDW gap size is not much affected, but the gap center shifts up. (e) When $\Delta_S$ and $\Delta_{SDW}$ are both moderately reduced, the effective SDW gap is both reduced in size and shifted up in energy, best matching the experimental observations.

**Fig. 4: Temperature-dependence of the SDW and SC gaps**

(a) EDCs divided by Fermi-Dirac function convolved with a Gaussian representing instrumental resolution taken across the SDW gap (integrated EDCs within a $\Delta k_x = 0.05\pi/a$ momentum range centered on the SDW gap for each temperature) as in Fig. 2(c) from above $T_S/T_{SDW}$ to below $T_C$. Each EDC is fitted with two Lorentzians and a linear background, with the fit shown by the black curve. Diamonds indicate the peak positions above and below the SDW gap, where the error bars indicate the standard deviations from the fits. Shaded region indicates the effective SDW gap as a function of temperature. (b) Symmetrized EDCs at the $k_F$ point of the band crossing as in Fig. 2c, for temperatures across $T_C$. Diamonds indicate lower branch of the SDW gap, circles indicate superconducting peaks. Black lines are symmetrized fits to the symmetrized EDCs with the Norman model[33] for the SC gap, a Lorentzian for the SDW peak, and a second order polynomial for the background. (c) The extracted effective SDW gap size for the undoped (black) and UD compound (red), along with the extracted SC gap size for the UD compound (blue), as a function of temperature. The error bars are from the properly combined standard deviations of the fitted positions of the two peaks in each EDC marking the upper and lower edges of the SDW gaps.

**Figure 1**

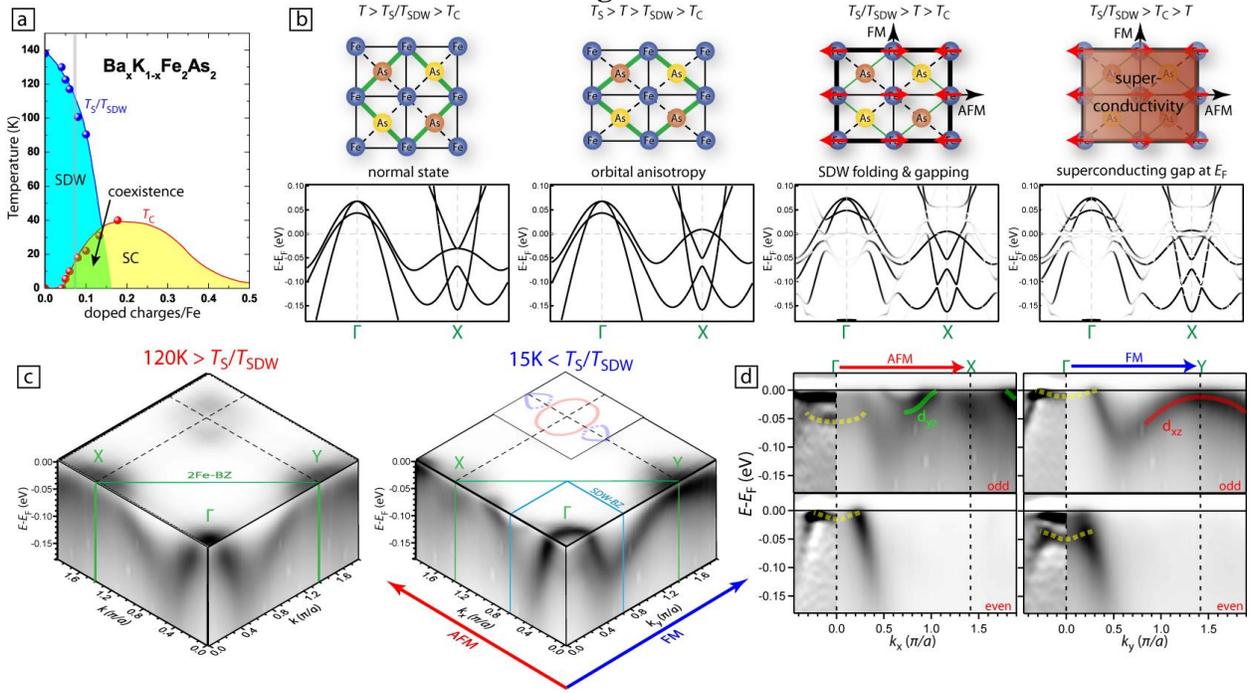

**Figure 2**

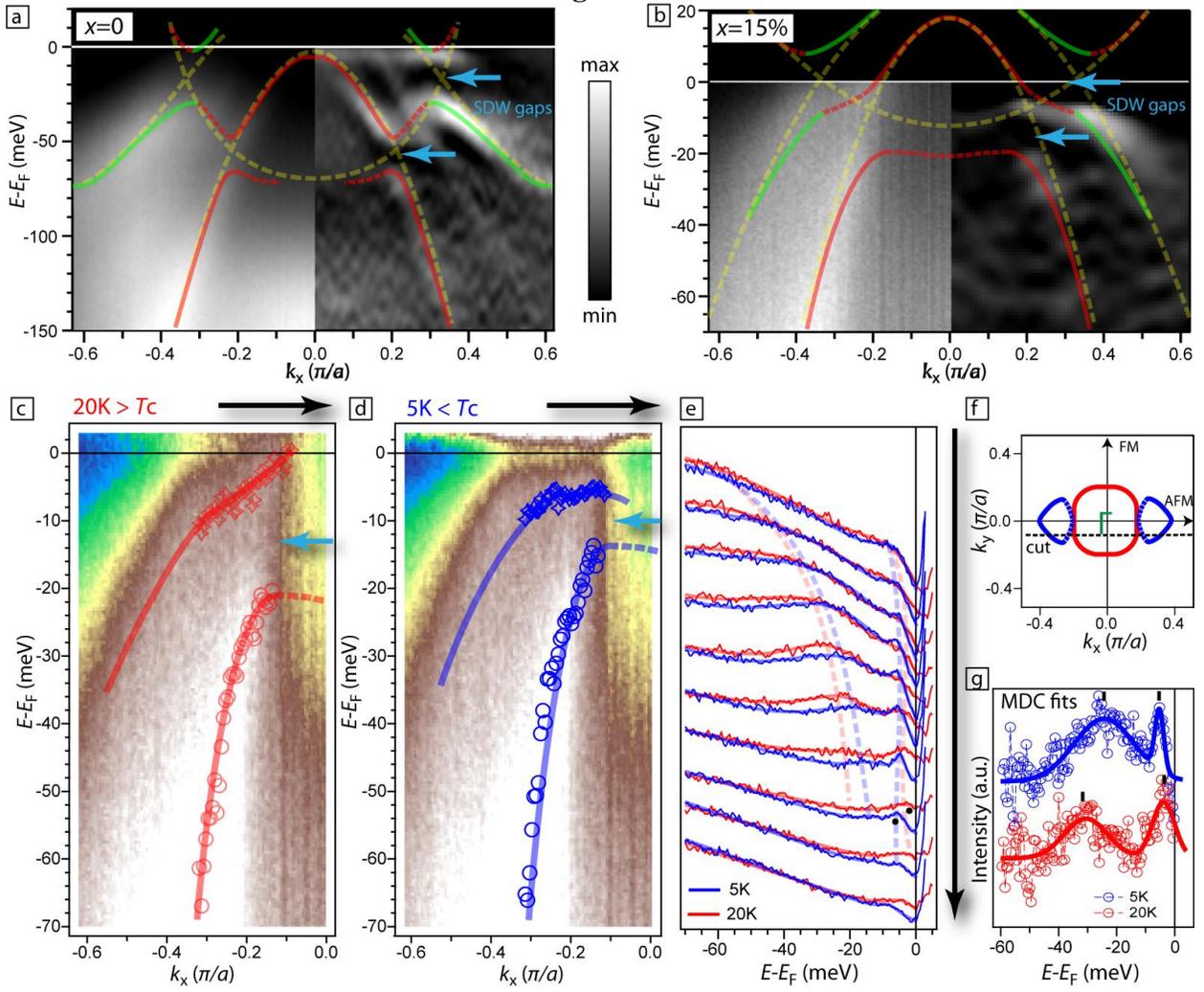

**Figure 3**

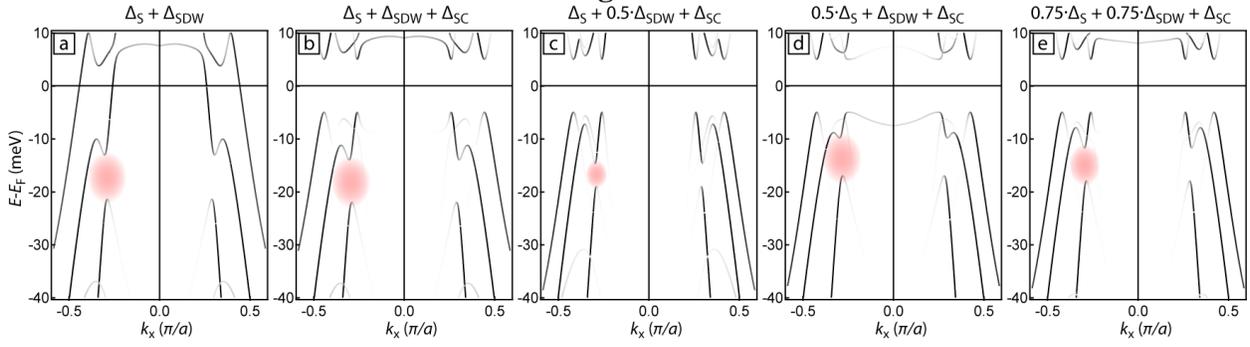

**Figure 4**

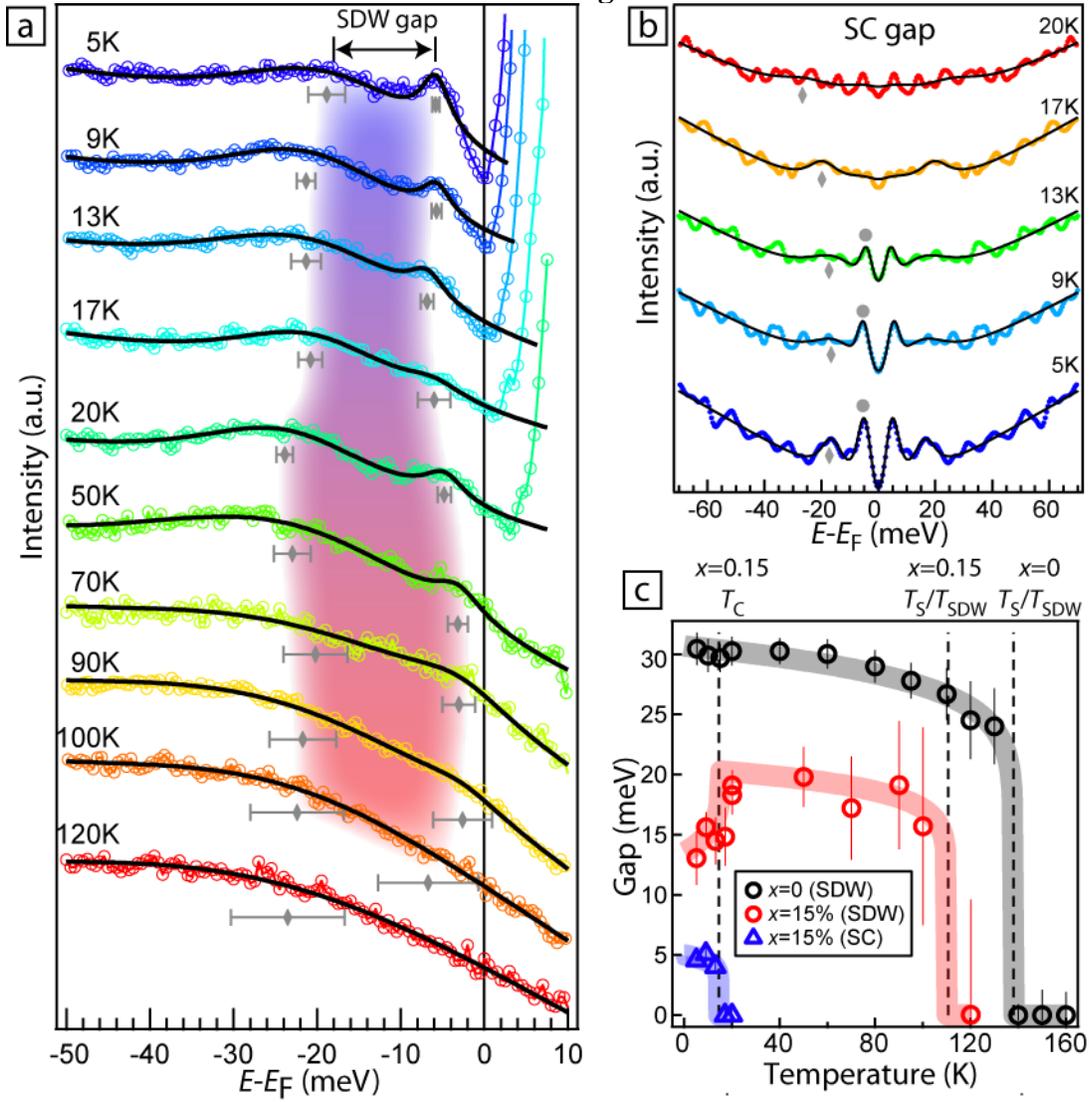

# Supplementary Figures

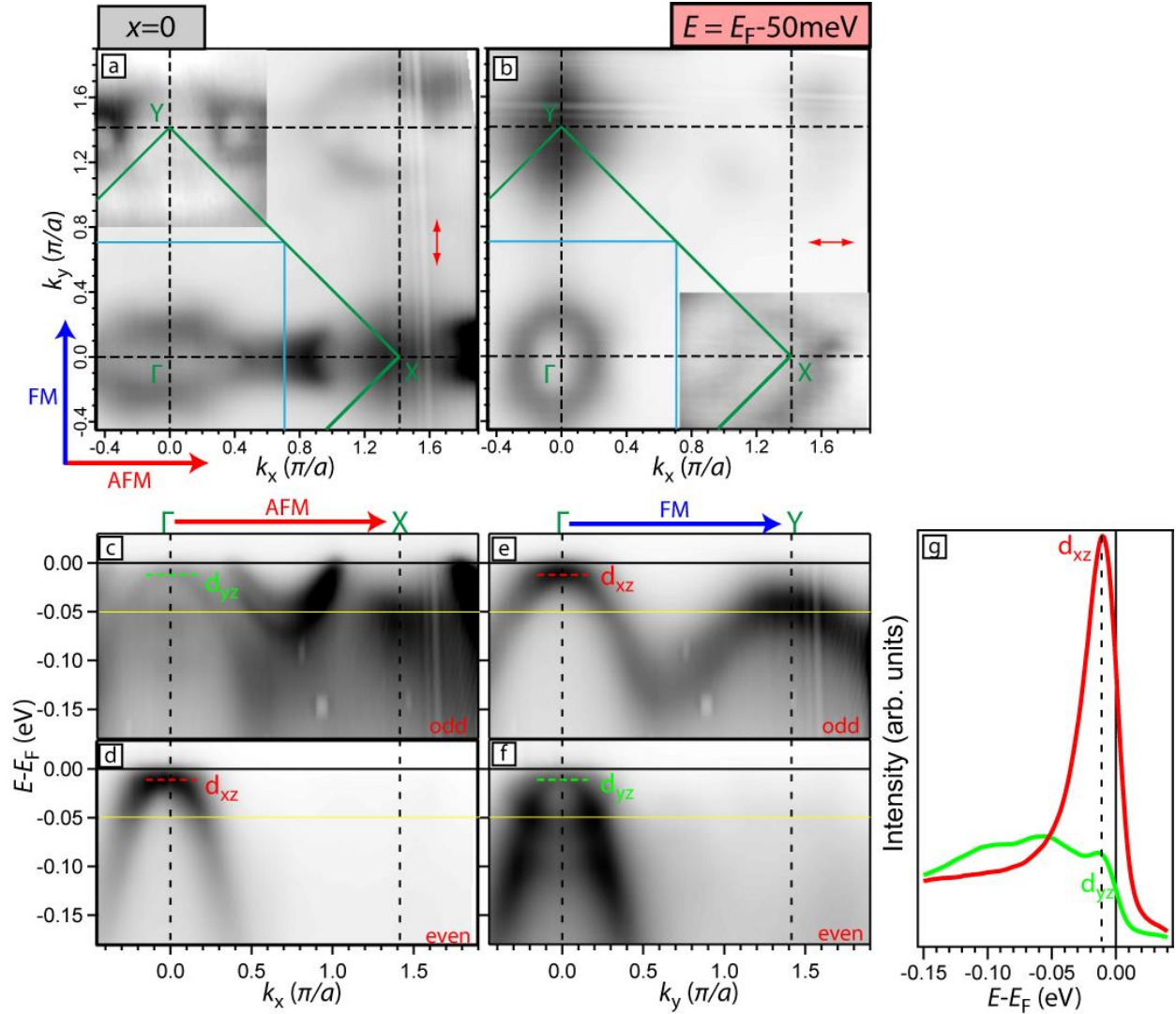

**Supplementary Figure 1: Electronic structure of undoped BaFe$_2$As$_2$**

(a)-(b) Constant energy mapping of undoped BaFe$_2$As$_2$ at 50meV below $E_F$ measured under different linear polarizations as labeled in red. Green lines mark the boundaries of the Brillouin zone (BZ) corresponding to the 2-Fe unit cell, while the blue one is that for the SDW BZ. (c)-(d) Band dispersions measured along the Γ-X AFM high symmetry direction, with light polarizations denoted in each panel. (e)-(f) Band dispersions measured along the Γ-Y FM high symmetry direction, with light polarizations denoted in each panel. Yellow lines mark 50meV below $E_F$. Dotted lines mark the energy positions of the hole band tops (red:$d_{xz}$, green: $d_{yz}$) at $\Gamma$. (g) Energy distribution curves (EDCs) taken at the $\Gamma$ point from panel (c) in green and (e) in red. The dotted line marks the energy position of the $d_{xz}$ and $d_{yz}$ band tops. All measurements were performed at 10K with 47.5eV photons, corresponding to $k_z = 0$.

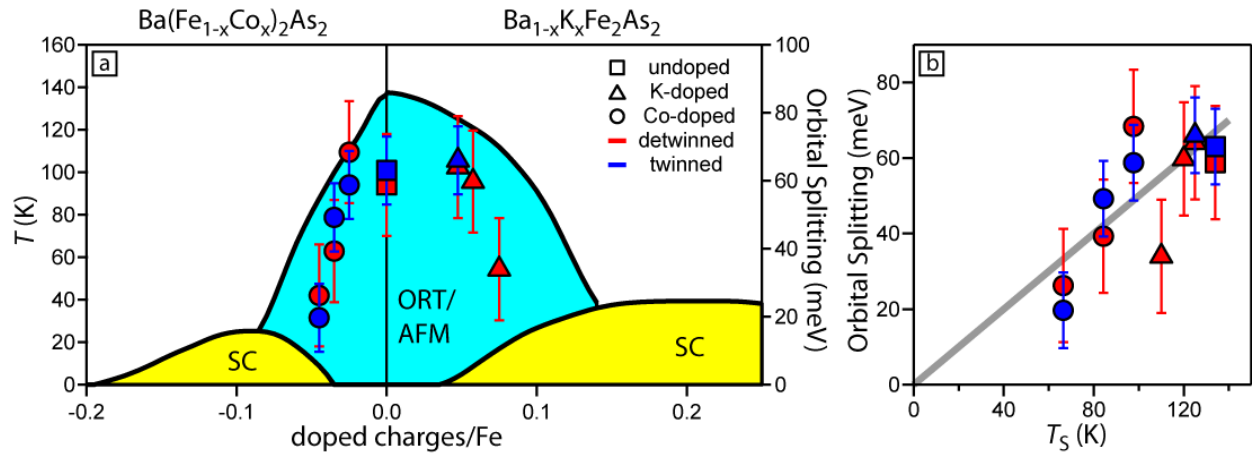

**Supplementary Figure 2: Doping dependence of orbital anisotropy**

(a) Orbital anisotropy measured as the energy difference between the $d_{yz}$ band along $\Gamma$-X and $d_{xz}$ band along $\Gamma$-Y at $k=0.9\pi/a$, for undoped (square), Co-doped[1] (circles) and K-doped (triangles) compounds, plotted on the phase diagram[2]. Both results from detwinned crystals (red) and twinned crystals (blue) are shown. (b) All the data points from (a) plotted against the structural transition temperature of each measured compound. The gray line is a guide to eye with a slope of 0.5meV/K. The legend is the same as that for (a). All measurements were taken at 10K with 47.5eV photons, corresponding to $k_z = 0$.

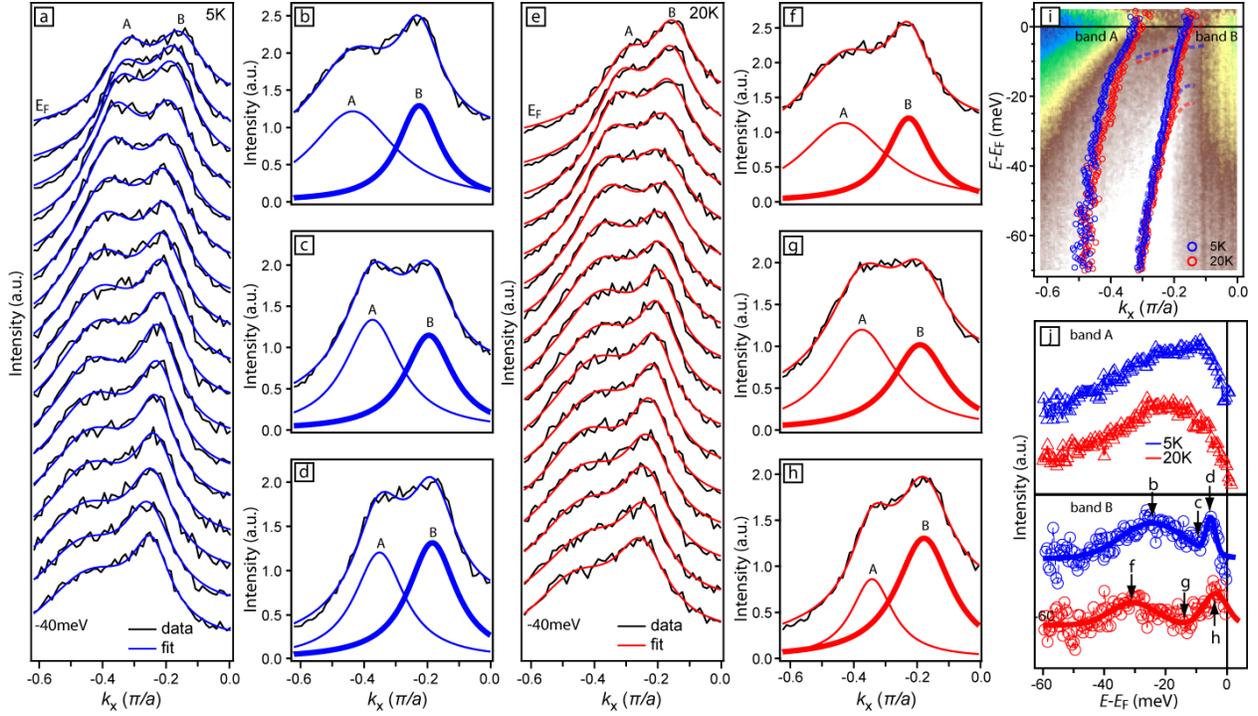

**Supplementary Figure 3: MDC fittings**

(a) Momentum distribution curves (MDCs) between $E_F$ and -40meV of Fig. 2d taken at 5K, with the fits of two Lorentzians and a linear background overlaid in blue. (b)-(d) Examples of the fits with the two Lorentzians at the energy positions labeled in (j). (e)-(h) Same as (a)-(d) but for measurements at 20K. (i) Fitted positions of the two Lorentzians (labeled A and B) for the two temperatures overlaid on spectral image taken at 20K. Dotted lines are the fitted band positions determined using EDCs as reproduced from Fig. 2c-d of the main text. (j) Fitted Lorentzian intensities as a function of energy for band A (upper) and band B (lower) for both temperatures.

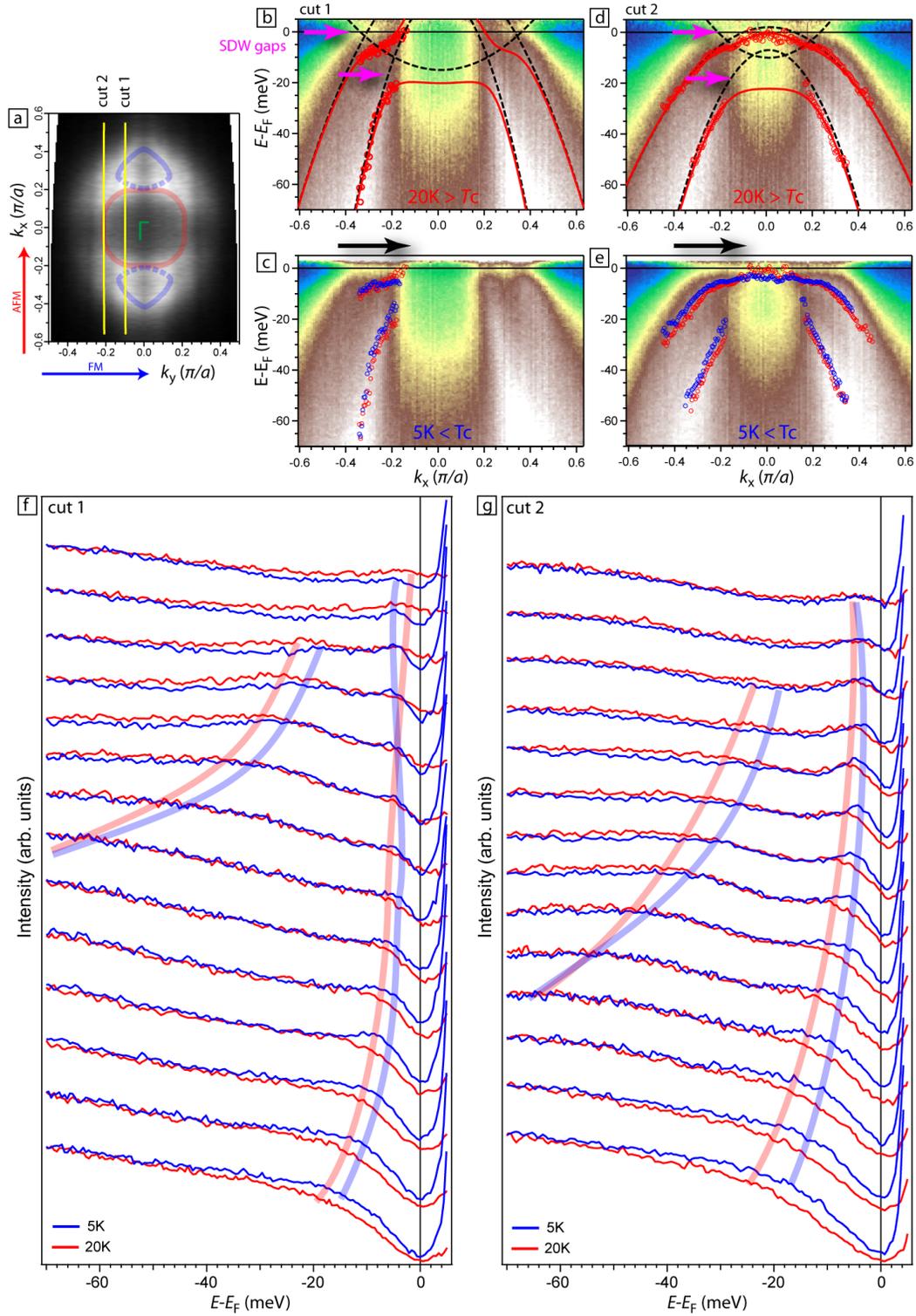

**Supplementary Figure 4: Momentum-dependence of the SDW-SC competition in Ba$_{0.85}$K$_{0.15}$Fe$_2$As$_2$**

(a) Fermi surface mapping at 5K around the BZ center. The positions of the two cuts are shown. (b) Band dispersion measured along cut1 at 20K>$T_C$, same as the one in the main text. Fitted band positions are shown by red circles. Schematic bands are shown in red lines, which are results of an SDW folded electron band and two hole bands (marked in black dotted lines). (c) Cut1 measured at 5K<$T_C$. Band positions for 20K>$T_C$ are overlaid in red circles for comparison with that measured for 5K<$T_C$, marked by blue circles. (d)-(e) Same as that in (b)-(c), but for cut2. (f) Comparison of energy distribution curves (EDCs) along cut1 in the momentum range marked by the black arrow in (c) for 5K and 20K. The dispersions in this momentum range are highlighted by following the peaks in the EDCs, marked by thick fainted blue (5K) and red (20K) lines. (g) Same as (f) but for cut2.

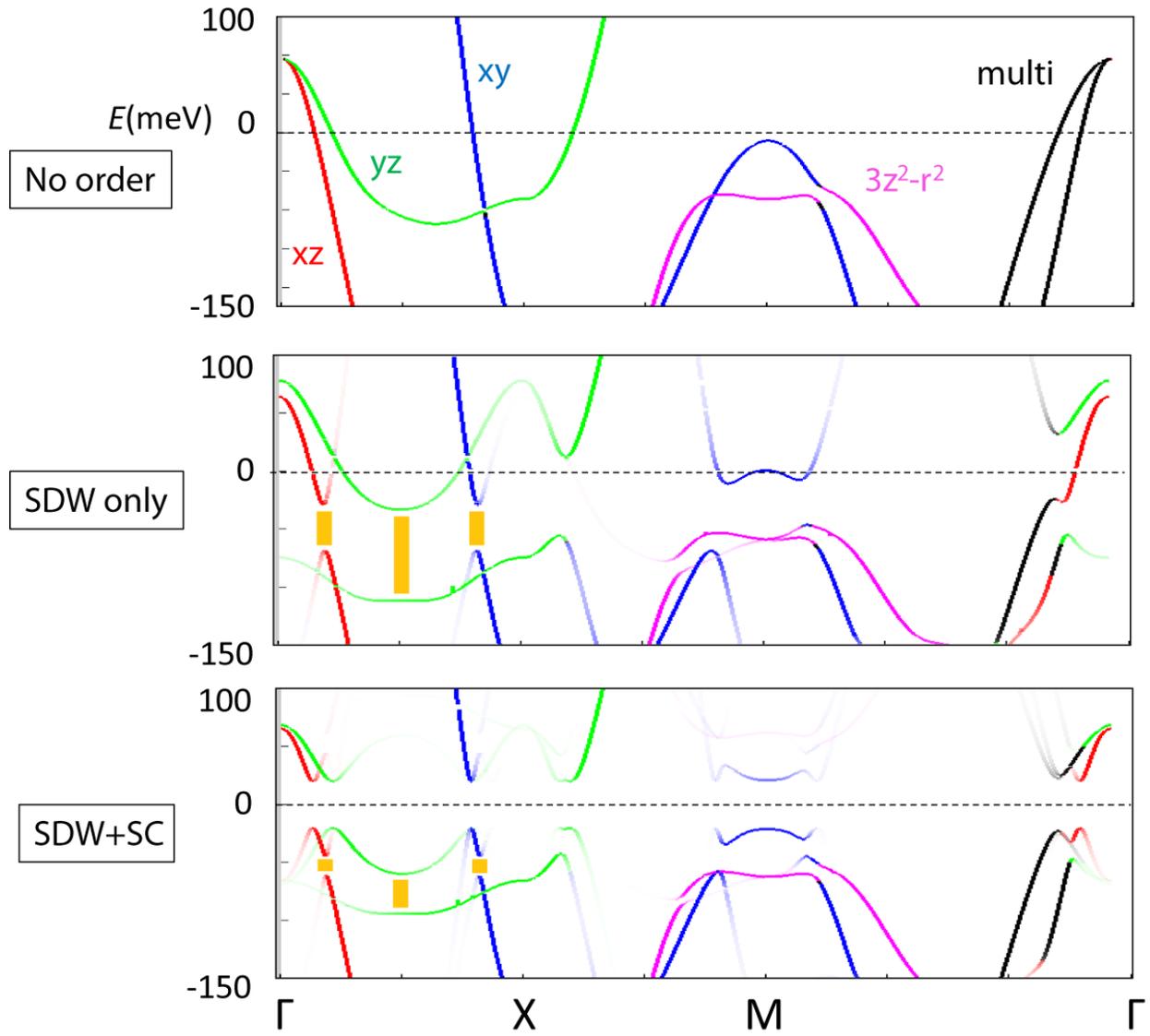

**Supplementary Figure 5: Band calculations showing SDW-SC competition**

Bands along high symmetry cuts at zero temperature compared for several cases of symmetry breaking. Band opacity displays spectral weight. Bands are colored as indicated according to the orbital with more than 50% of spectral content, or black if no orbitals dominate. In the bottom panel, SC gaps are generally distinguished by being centered on the Fermi level, unlike SDW gaps. The bars indicate SDW gap magnitudes, which shrink with the inclusion of SC.

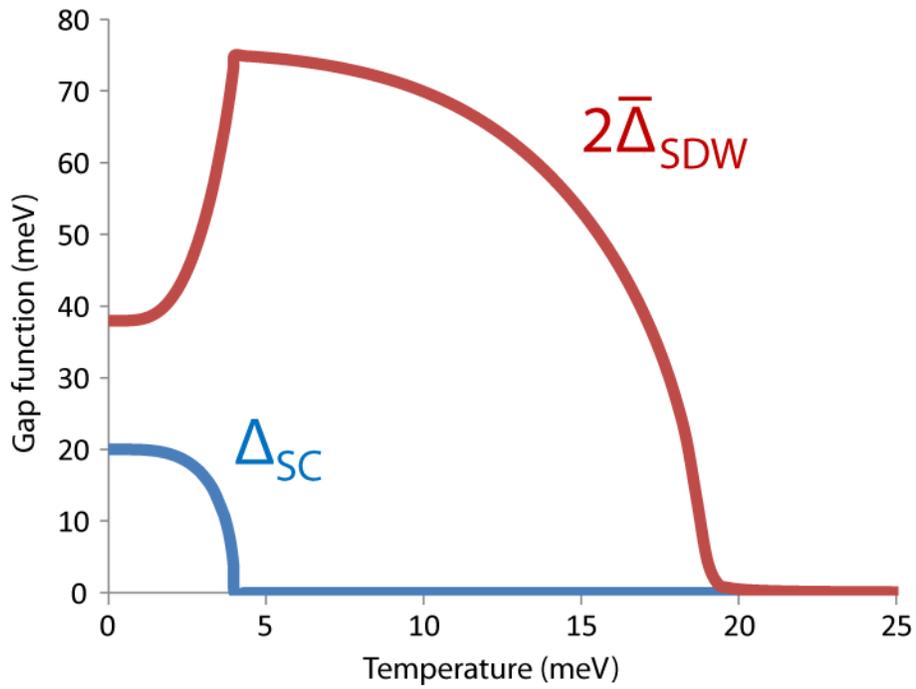

**Supplementary Figure 6: Calculated competition of SC and SDW order parameters**

Temperature evolution of the SC and SDW gaps, displaying coexistence and competition between these phases.

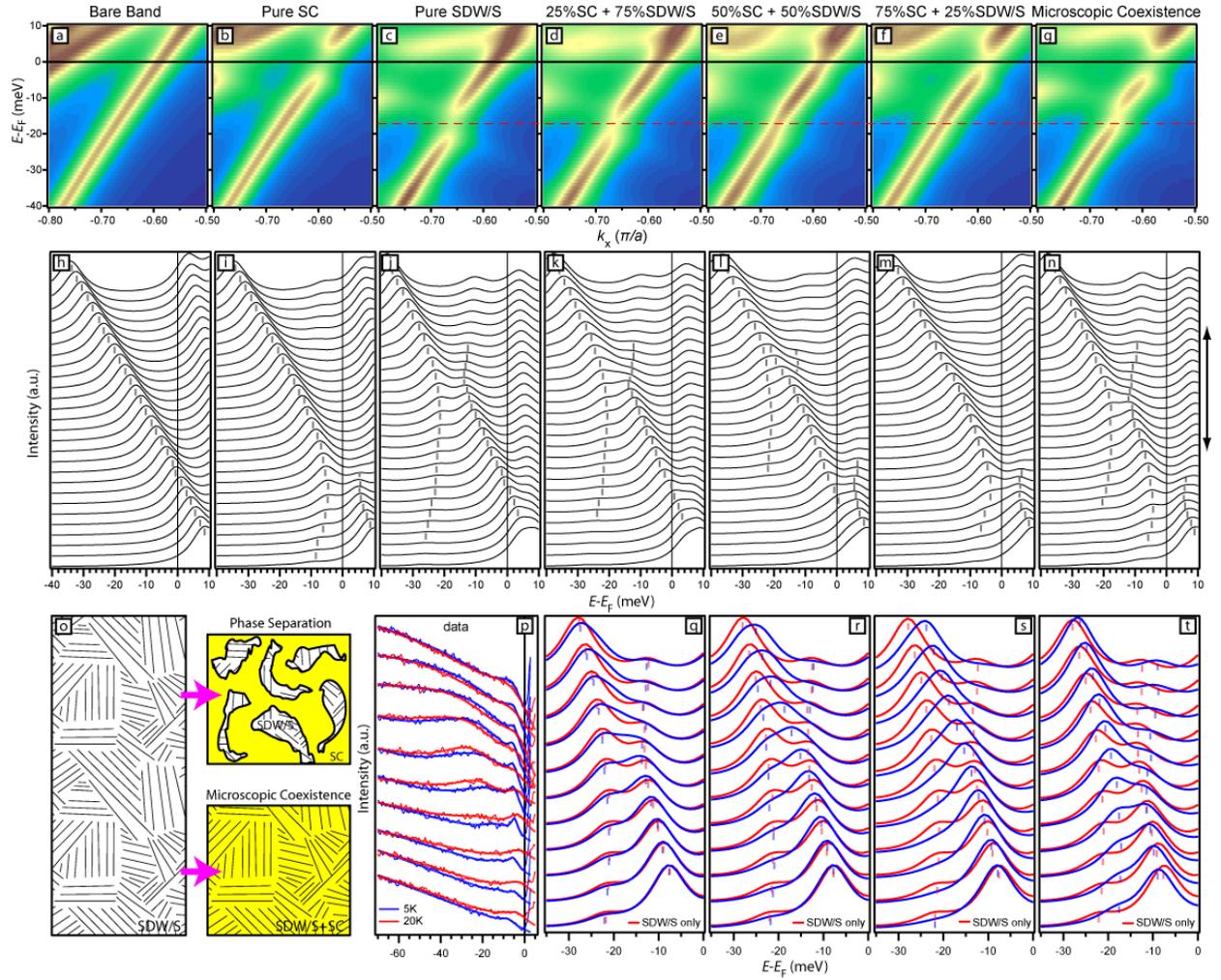

**Supplementary Figure 7: Simulation for phase separation versus microscopic coexistence**
Simulated ARPES spectra image along the same cut as that in Fig. 3 of the main text, with artificial line-width broadening, $\Gamma(\omega)=\alpha+\beta*\omega^2$, ($\alpha$ = 5meV, $\beta$ = 5meV$^{-1}$), for (a) bare band, (b) pure SC, (c) pure SDW and S orders, (d) mixed signal of SC regions and SDW/S regions with 25% SC volume fraction, (e) 50% SC volume fraction, (f) 75% SC volume fraction, (g) microscopic coexistence with reduced SDW and S order parameters as in the case of Fig. 3e. The dotted line is drawn at the center of the SDW gap in the pure SDW/S case for comparison with the various scenarios. (h)-(n) Energy distribution curves (EDCs) for all the cases in the upper panels, with peaks marked out. (o) Schematic showing the scenarios of phase separation and microscopic coexistence. (p) Reproduction of the data from Fig. 2e for comparison. (q)-(t) EDCs across the SDW gap in the range marked by the arrow in (n) for the various cases (blue) compared with the case of pure SDW/S (red), where the prominent peaks are marked.

# Supplementary Notes

**Supplementary Note 1: Comparison of Ba$_{0.85}$K$_{0.15}$Fe$_2$As$_2$ to Ba(Fe$_{1-x}$Co$_x$)$_2$As$_2$**

The electronic structure of underdoped (UD) K-Ba122 is very similar to that of undoped BaFe$_2$As$_2$, except for a shift of the chemical potential due to hole doping. Supplementary Figure 1 shows the electronic structure of undoped BaFe$_2$As$_2$ taken in the orthorhombic SDW state. Comparing the band dispersions of undoped and UD compounds, we see that both the orbital-dependent band shift and band folding are similar in the two compounds. In particular, the direction of the orbital-dependent band shift of the two compounds is the same, in contrast to the reported reversal of resistivity anisotropy on the K-doped side as compared to the undoped and electron-doped side[3], suggesting that the dominant cause of the resistivity anisotropy is likely to be related to changes in scattering, rather than the band dispersions. While it is not merely a rigid chemical potential shift between the undoped and UD compound, we see that lowering the chemical potential by an energy scale of ~50meV in the undoped compound reveals electronic structure that is largely similar to that of the $x$=0.15 UD compound, as seen in both the constant energy mapping and the band dispersions (Supplementary Figure 1).

In addition, we would like to briefly comment on the momentum-dependence of the orbital anisotropy. In the undoped BaFe$_2$As$_2$, at $k_z$=0, the band tops of the hole bands at Γ are still below $E_F$, and hence experimentally accessible (Supplementary Figure 1c-f). In the polarization dependence study here, we can resolve the d$_{xz}$ hole band in Supplementary Figure 1d and e, and d$_{yz}$ hole band in Supplementary Figure 1c and f, from which we see that the hole band tops are not much shifted between d$_{xz}$ and d$_{yz}$. This can also be seen in the EDC taken at the Γ point in the two different polarizations (Supplementary Figure 1g). However at the X/Y points, we see that the $d_{yz}$ hole-like band top is shifted to above $E_F$ while the $d_{xz}$ hole-like band remains below $E_F$, showing the orbital anisotropy. Since we do not expect qualitative differences between undoped and doped compounds, we incorporated such a momentum-dependent orbital-dependent shift as our $\Delta_S$ order parameter, with zero shift at $\Gamma$ and maximum shift at X/Y.

In addition to the $T_C$=14.5K sample, we have also measured the electronic structure of two other K-doped samples of smaller doping, with $T_C$=2K and 8K, both of which show the orbital anisotropy in the orthorhombic state. To quantitatively compare the orbital anisotropy of the hole doped (K-doped) compounds with that of electron doped (Co-doped) compounds, we have plotted the orbital energy splitting of the $d_{xz}$ and $d_{yz}$ bands for both kinds of doping together, all taken at the same momentum point (Supplementary Figure 2a). Here we see that the magnitude of the energy splitting roughly follows the monotonic trend of the structural transition, $T_S$, for both kinds of doping. To see this more clearly, we plot this energy splitting against the $T_S$ for each compound measured (Supplementary Figure 2b), which shows a rough linear relationship regardless of the type of doping, suggesting the universality of the orbital anisotropy in iron pnictides. Finally, we see that the full magnitude of the orbital anisotropy developed at the lowest temperature is the same for both unstressed twinned crystals and stressed detwinned crystals.

**Supplementary Note 2: MDC fittings showing reduction in SDW gap**

The second way of observing the SDW gap suppression in the superconducting state as mentioned in the main text is to compare the momentum distribution curves (MDCs) of the 20K and 5K spectra in Fig. 2c-d in the main text. Representative MDCs between the energy range of $E_F$ and -40meV for below and above $T_C$ are shown in Supplementary Figure 3a and e, respectively. These MDCs show two peak features, whose positions follow the two bands (labeled A and B). We fit the MDCs with two Lorentzians and a linear background. The fits are overlaid on the data in Supplementary Figure 3a and e. The fitted positions of the Lorentzians for the two bands for both temperatures are marked in Supplementary Figure 3i. Because these two hole bands are steep and that the SDW gap opens in energy, the position of the bands fitted by MDCs would not show the SDW gap, but the original ungapped bands. However, the presence of the SDW gap could be seen in the suppression of the fitted intensity of the Lorentzians (Supplementary Figure 3j). For band A, the SDW gap straddles $E_F$, as marked in Fig. 2b of the main text. Hence the fitted intensity shows a monotonic behavior until it dips approaching $E_F$. For band B, the SDW gap is below $E_F$, seen in the dip in the intensity between the two local maxima. The fitting details of the MDCs at the intensity local maxima and minima are shown in Supplementary Figure 3b-d for 5K, and Supplementary Figure 3f-h for 20K. From 20K to 5K, we see that this energy separation between the local maxima decreases, signaling that the SDW gap shrinks going below $T_C$.

**Supplementary Note 3: Momentum-dependence of SDW-SC competition**

Reduction of the SDW order below $T_C$ is not limited to a particular occurrence of the SDW gap. In this section we show the competition in a larger momentum range. Supplementary Figure 4b-c shows the same cut as the one in the main text. There are two SDW gaps on this cut, as discussed in the main text (Fig. 2b). Besides the lower one around -15meV discussed in the main text, we see that the one straddling $E_F$ also changes across $T_C$. In particular, we can see from the EDCs that the lower branch of this SDW gap shifts towards $E_F$, signaling the reduction of the SDW gap.

Supplementary Figure 4d-e shows another cut further away from the Γ point. Along this cut, in the SDW state, there is no Fermi crossing due to the SDW gapping between a folded electron band and two hole bands whose band tops are already below $E_F$ (Supplementary Figure 4d). Here again, there are two momenta where SDW gaps open, as indicated by the magenta arrows. For the lower SDW gap, the lower branch as seen in the EDC peaks is not strong, but a shift in central weight towards $E_F$ can be observed below $T_C$, similar to the behavior in cut1. For the higher SDW gap, only the lower branch can be observed, which again shifts towards $E_F$ below $T_C$, indicating a shrinking SDW gap.

The two cuts shown here are examples where significant SDW gaps can be observed in the band structure. All of the SDW gaps observed here show a reduction at the onset of SC. In momentum space where no SDW gap opens, such as near Γ exactly along the high symmetry line Γ-X, little change occurs across $T_C$ besides the opening of a SC gap (not shown).

**Supplementary Note 4: Hartree-Fock calculation for SDW-SC competition**

We have performed five-orbital Hartree-Fock calculations that support interpreting our observations as phase coexistence and competition between SC and SDW. Here we do not fit the bands to ARPES but rather determine similar bands from theory, to derive qualitative results. The starting point is a 10 × 10 mean field Hamiltonian matrix in spin sector $\sigma$ with only SDW order. Displayed in 5 x 5 blocks, each in the iron d-orbital basis,

$$H'_{MF}(\mathbf{k},\sigma) = \begin{pmatrix} H_0(\mathbf{k}) & \sigma M \\ \sigma M & H_0(\mathbf{k}+\mathbf{Q}) \end{pmatrix}. \quad (1)$$

$H_0(\mathbf{k})$ is the kinetic energy matrix at momentum $\mathbf{k}$. It is determined from a tight-binding fit to the density functional theory band structure of LaFeAsO[4], which is qualitatively similar to $Ba_{1-x}K_xFe_2As_2$ at low energies and dopings[5], and does not change the qualitative result of this study. For the magnetic part, we chose the nesting vector $\mathbf{Q} = (\pi, 0)$, and $M$ is the SDW gap matrix amongst orbitals. Self-consistent calculations show that $M$ is diagonal except for small off-diagonal terms between the two $e_g$ orbitals[6]. Therefore, bands that cross upon magnetic backfolding are only gapped if they have essentially similar orbital content. Diagonalizing Eq. 1 yields a matrix in SDW band basis, which we denote $\widehat{H}'_{MF}(\mathbf{k},\sigma)$. We renormalize bands by a factor of 3 to better match the bandwidth observed with photoemission[1].

This obtained matrix in band basis is then used to construct the $20 \times 20$ mean field Hamiltonian with SC and SDW. Displayed in $10 \times 10$ blocks, each in the SDW band basis,

$$H_{MF}(\mathbf{k}) = \begin{pmatrix} \widehat{H}'_{MF}(\mathbf{k},\uparrow) & \widehat{\Delta} \\ \widehat{\Delta} & -\widehat{H}'_{MF}(-\mathbf{k},\downarrow) \end{pmatrix}. \quad (2)$$

We chose the SC gap matrix $\widehat{\Delta}$ to be diagonal (i.e. intra-band), with all diagonal entries equal, since we do not observe strong band gap variations in experiment. The magnetic moment $M$ is determined by self-consistent Hartree-Fock on multi-orbital electron-electron interactions

$$H_{int} = U \sum_{i,r} n_{i,r\uparrow} n_{i,r\downarrow} + U' \sum_{\substack{i,r,s<r \\ \sigma\sigma'}} n_{i,r\sigma} n_{i,s\sigma'} + J \sum_{\substack{i,r,s<r \\ \sigma\sigma'}} c^\dagger_{i,r\sigma} c^\dagger_{i,s\sigma'} c_{i,r\sigma'} c_{i,s\sigma} + J' \sum_{i,r,s<r} (c^\dagger_{i,r\uparrow} c^\dagger_{i,r\downarrow} c_{i,s\downarrow} c_{i,s\uparrow} + h.c.)$$

$$(3)$$

in which $c^{\dagger}_{i,r\sigma}$ creates an electron at site *i* with $\sigma$ denoting its spin and *r* (or *s*) its orbital. The particle densities are $n_{i,r} = n_{i,r\uparrow} + n_{i,r\downarrow}$. Spin rotational invariance requires $J' = J$ and in the absence of exchange anisotropy $U' = U - 2J$ holds for each pair of interacting orbitals, which we take to be independent of band indices. We apply these simplifications to our Hamiltonian and take *U* = 433 meV and $J = U/4$, which stabilizes a ($\pi$,0) SDW with a small magnetic moment of 0.4. (More details can be found in a similar calculation in Ref. 6.) The self-consistent calculation is performed in the presence of fixed isotropic superconducting gap $\widehat{\Delta}$ and filling *n* = 6. Doping effects do not qualitatively affect the analysis. A BCS temperature dependence was assumed for the phenomenological superconducting gap[7]. The transition temperature was chosen to be $T_{SC}$ = 4 meV and zero temperature gap 20 meV. These choices are (1/5)$T_{SDW}$ and roughly half of the average zero temperature SDW gap in the absence of SC, respectively, and follow experimentally observed ratios.

We find that in the presence of SC, all SDW gaps coexist with SC and they shrink due to phase competition. Supplementary Figure 5 compares the *T*=0 bands for SDW alone to those of SDW in the presence of SC. With the opening of SC gaps, all SDW gaps continue to exist but shrink, as the bars highlight. The SDW gaps shrink due to two effects from SC: First, all non-zero entries of the self-consistently determined gap function $M$ shrink to approximately half of their value, as SC suppresses the magnetic order. Second, SC gaps open around the Fermi level, and all bands that are near the Fermi level are similarly repelled towards higher binding energies. The SC gaps remove quasiparticles around the Fermi surface that would otherwise be available to form the SDW. The calculated band behavior of SC and SDW gaps is qualitatively consistent with experimental observations.

This coexistence and competition follows the experimentally observed temperature dependence. Supplementary Figure 6 shows the evolution of SC and SDW gap functions as temperature decreases. The SC gap $\Delta_{SC}$ is the value of all diagonal entries of $\widehat{\Delta}$, and the SDW gap $\overline{\Delta}_{SDW}$ is the average value of the diagonal entries of $M$. Averaging is sufficient since all gaps shrink by the same fraction. As SC increases, SDW decreases monotonically in a manner similar to the observed gap reduction (Fig. 4c). We compare $\Delta_{SC}$ to $2\overline{\Delta}_{SDW}$ in correspondence with Fig. 4c of the main text: with the SDW gap as the energy difference between the gap edges, and the SC gap taken as the energy difference between the lower gap edge and the Fermi level. These results are consistent with our observations, supporting the interpretation of phase competition between SC and SDW.

**Supplementary Note 5: Distinguishing phase separation and microscopic coexistence**

There are two distinct ways in which the SDW order and superconductivity could coexist: phase separation and microscopic coexistence. For a phase separated scenario, the SDW order and superconducting (SC) order exist in spatially separated regions in the material, and are not finite simultaneously at any spatial point. For a microscopic coexisting scenario, the two order parameters are finite simultaneously at the same spatial point, and extend over the whole compound (Supplementary Figure 7o). Since ARPES is a spatially averaging technique, the two scenarios could both lead to the simultaneous observation of the signatures of the SDW order and SC order in the same spectra. However, we can distinguish the two scenarios by simulating the ARPES spectral under each scenario and comparing with our data. We take the tight-binding model used for Fig. 3, and add a finite line-width broadening, $\Gamma(\omega)=\alpha+\beta*\omega^2$, where $\alpha = 5$meV, $\beta$

= 5meV$^{-1}$, to simulate the ARPES spectral. All order parameters used here are the same as the ones for Fig. 3.

Supplementary Figure 7a shows the simulated bare ARPES spectra in the region of interest, with no order of any kind. A bare band is seen crossing $E_F$, as also seen in the EDC plot in Supplementary Figure 7h. First, we simulate the case of phase separation. In Supplementary Figure 7b and c, we simulate the cases of pure superconducting order and pure SDW/S orders, respectively. In the case of pure SC, a particle-hole symmetric gap opens at $E_F$, as seen in the EDC plot (Supplementary Figure 7i). In the case of pure SDW/S, an SDW gap opens around -18meV. This case mimics the compound deep in the orthogonal SDW state but above $T_C$. When SC sets in, for a phase separated scenario, the compound would develop regions of pure SC and hence decrease the volume fraction of pure SDW/S orders without reducing the order parameter magnitudes within each domain. For an ARPES experiment where the beam size is much larger than the typical domain size, the signal would have contributions of pure SC and pure SDW/S regions. To simulate this, we add the spectra of pure SC and pure SDW/S in different proportions to imitate cases of differing SC volume fraction (Supplementary Figure 7d-f), along with the EDC plots (Supplementary Figure 7k-m). For the 25% SC volume fraction case, the SC gap at $E_F$ is too weak to be seen, and the SDW gap is little affected (Supplementary Figure 7q). For the 75% SC volume fraction case, one can no longer detect an SDW gap as the EDCs only show a single peak due to the dominance of the SC region signal (Supplementary Figure 7s). We focus on the 50% SC volume fraction case, where both order signatures are most detectable (Supplementary Figure 7e and l). There are three observations to note. Firstly, the SC gap becomes slightly asymmetric about $E_F$, due to the mixed signal from the SDW regions. Secondly, the SDW gap now has an extra peak that fills inside the original SDW gap, due to the mixing of the signal from the SC regions where there is no SDW gap. Thirdly and most importantly, the position of the SDW gap does not shift towards $E_F$ as compared to the case of pure SDW/S order. All three of these observations are inconsistent with the actual experimental observation (Supplementary Figure 7p). Next, we simulate the case of microscopic coexistence by introducing both SC and SDW/S order parameters to the same band structure, and reducing the $\Delta_{SDW}$ and $\Delta_S$ to 0.75 of the full values, corresponding to the case in Fig. 3e. The result is in Supplementary Figure 7g. Here we see in the EDC plot (Supplementary Figure 7n) of a particle-hole symmetric SC gap, and an SDW gap that is both reduced in size and shifted towards $E_F$ when compared to the case of pure SDW/S order (Supplementary Figure 7t).

Hence, we conclude that our experimental observations are not consistent with the phase separated scenario mentioned above and are more consistent with a microscopic coexistence scenario. However, we do note that the case of phase separation simulated above that we have excluded is the most generic one in which we assume no interaction between the SC and SDW/S regions, and that the SDW/S domain sizes are much larger than the SDW/S coherence length such that the decrease in volume fraction does not affect the order parameters directly.

## Supplementary References